\documentclass[seceq]{ptptex}

\usepackage{graphicx}

\newcommand{\HII}{H{\sc ii} }
\newcommand{\HI}{H{\sc i} }

\newcommand{\cm}{21$\,$cm }

\newcommand{\nc}{\newcommand}

\nc{\beq}{\begin{equation}}
\nc{\eeq}{\end{equation}}
\nc{\bea}{\begin{eqnarray}}
\nc{\eea}{\end{eqnarray}}
\nc{\n}{\nonumber \\}

\title{
The Dark Ages of the Universe and Hydrogen Reionization
}

\author{
Aravind \textsc{Natarajan}$^{1}$\footnote{E-mail: aravind@pitt.edu} and
Naoki \textsc{Yoshida}$^{2,3}$
}

\inst{
  $^{1}$Department of Physics and Astronomy \& Pittsburgh Particle physics, Astrophysics and Cosmology Center, University of Pittsburgh, 100 Allen Hall, 3941 O'Hara Street, Pittsburgh, PA 15260, U.S.A. \\
$^{2}$Department of Physics, University of Tokyo, Bunkyo, Tokyo 113-0033, Japan\\
  $^{3}$Kavli Institute for the Physics and Mathematics of the Universe (WPI),\\
  University of Tokyo, Kashiwa, Chiba 277-8583, Japan\\
}

\abst{
One of the milestones in the cosmic history
is the formation of the first luminous objects and hydrogen reionization. 
The standard theory of cosmic structure formation predicts that 
the first generation of stars were born
about a few hundred million years after the Big Bang.
The dark universe was then lit up once again, and
eventually filled with ultra-violet photons emitted
from stars, galaxies, and quasars.
The exact epoch of the cosmic reionization 
and the details of the process, even the
dominant sources, are not known except the fact that 
the universe was reionized early on. 
Signatures of reionization are expected to be
imprinted in the cosmic microwave background radiation,
especially in its large-scale polarization.
Future CMB experiments, together with other probes such as
\HI \cm surveys, will provide rich information on the process of reionization.
We review recent studies on reionization. The implications
from available observations in a wide range of wavelengths 
are discussed.
Results from state-of-the-art computer simulations are presented.
Finally, we discuss prospects for exploring the first
few hundred million years of the cosmic history.
}

\begin{document}

\maketitle

\section{Introduction}
The quest for neutral hydrogen in the inter-galactic 
space has a long history
\cite{GunnPeterson}. Observations in the 1960's 
surprisingly showed that there is indeed little amount of 
neutral hydrogen in the inter-galactic medium (IGM).
It was then immediately proposed that the IGM itself  
is in a highly ionized state rather 
than being neutral.
A question then naturally followed: 
how was the IGM ionized? 
It was not until 2000 that the so-called Gunn-Peterson trough
was finally found in the spectra of distant quasars\cite{fan1}.
The observations reached an early epoch when cosmic 
reionization was being completed. Clearly, the inter-galactic gas 
had been indeed neutral but was ionized at an early epoch
by some sources of radiation or by some other physical mechanism.

A number of more recent observations suggest that the universe was
reionized early on, in the first several hundred million 
years. For example, CMB experiments provide information on the epoch
of reionization through the measurement of the total Thomson
optical depth. 
The scattering of CMB photons with free electrons is quantified 
by means of the optical depth:
\begin{eqnarray}
\tau &=& \int dt \, c \,   \sigma_{\rm T} \, n_{\rm e} 
=  \int \frac{dz}{H(z)(1+z)} c \, \sigma_{\rm T} \, n_{\rm e}(z) \\
&=& \frac{c \, \sigma_{\rm T} \left( \rho_{\rm crit} / h^2 \right )}{100 \; {\rm km/s/Mpc}} \; \frac{1}{m_{\rm N}} \; \frac{ \Omega_{\rm b} h^2}{\sqrt{\Omega_{\rm m}h^2}} \; 
\left( 1 - Y \right ) \\ 
&\times& \int dz \frac{ \left(1+z\right )^2}{\sqrt{\left(1+z\right)^3 + \left (\Omega_\Lambda/\Omega_{\rm m} \right )}} x_{\rm e}(z) \left [ 1 + \mu(z) \frac{Y}{4 \left(1-Y\right)} \right ]. \nonumber
\label{tau}
\end{eqnarray}
where $x_{\rm e}(z)$ is the ionization fraction, 
$c$ is the speed of light, $\sigma_{\rm T}$ is the Thomson cross section, $m_{\rm N}$ is the nucleon mass,
and $Y$ is the helium fraction.
The pristine IGM consists of hydrogen and helium. 
We set $\mu(z)$ = 0 if helium is neutral, 
1 if singly ionized, and 2 if doubly ionized. 
The large-scale polarization of the cosmic microwave 
background measured by the WMAP satellite suggests
that reionization -- release of free electrons --
began as early as $z \sim 10$.\cite{Hinshaw}
There are also other indirect probes, from the cosmic infrared
background to the distribution of star-forming galaxies
at $z > 6$. The ionized fraction, or alternatively the neutral fraction, 
of the IGM can be measured in multiple ways.\cite{Fan12}
Ultimately, all  such observations
must be explained as outcomes of a series of events 
that affected the ionization and thermal state of the IGM 
in the early cosmic history.

Theoretical studies on reionization naturally include the
formation of the first cosmic structures\cite{YoshidaPTEP}.
In this article, we first review recent progress in the theory of 
structure formation in the early universe. 
We then give an overview of probes of cosmic reionization
and the Dark Ages. We put our emphasis on the use of hydrogen
\cm emission and absorption to be observed in the currently
operating and future radio telescope arrays.
We conclude the present article by discussing the prospects
for direct and indirect observations of reionization.

\begin{figure}
\begin{center}
\scalebox{0.3}{\includegraphics{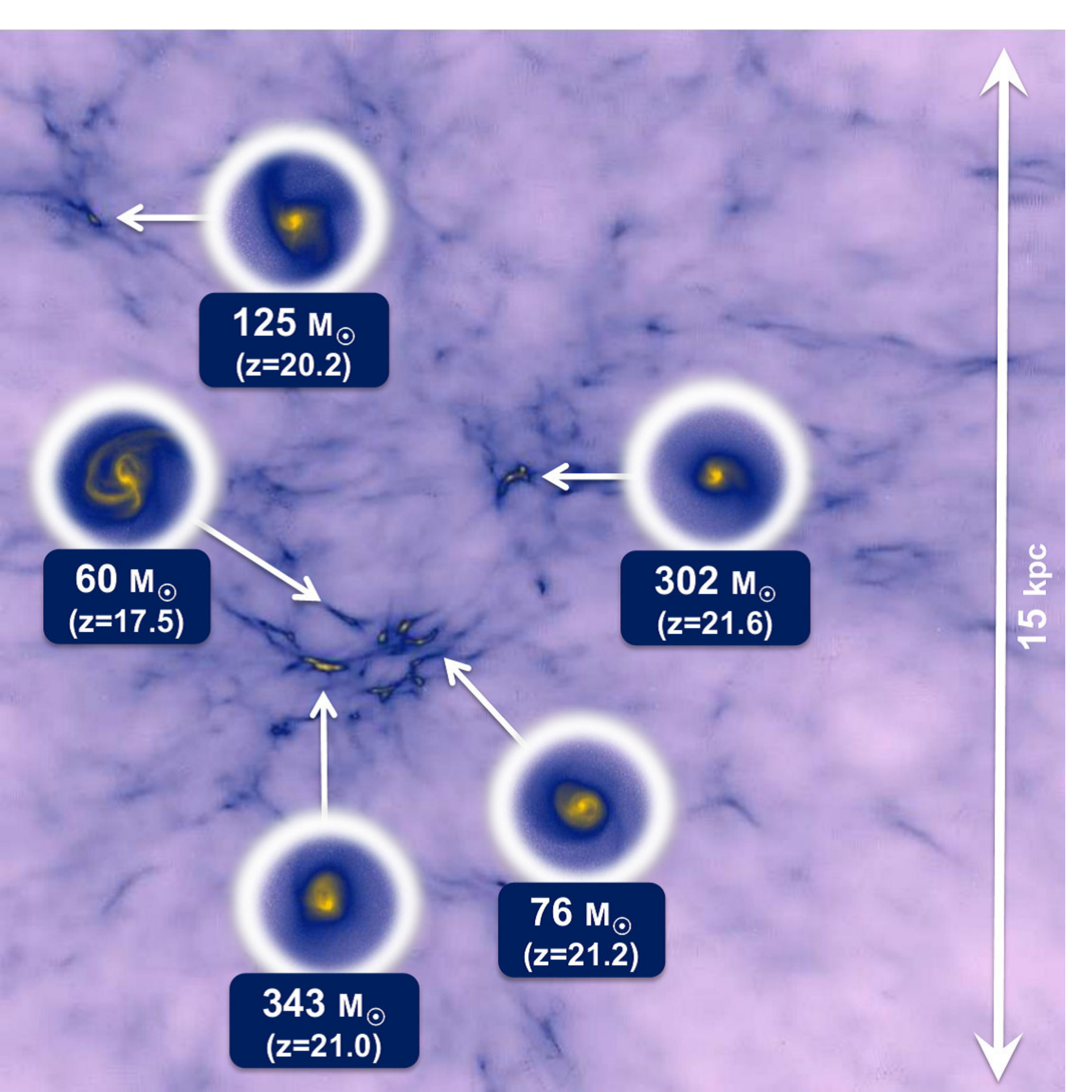}}
\caption{The matter distribution in the early universe. 
The plotted region is a cube of 15 kpc on a side.
First stellar nurseries are found at the knots of the filamentary
structure. The insets show the fine structure of the star-forming
regions. Also the final masses of the newly born stars
are indicated. From Ref.~\citen{Hirano13}.
\label{ny_fig1} }
\end{center}
\end{figure}

\section{Early structure formation}
We begin by describing  structure formation
in the standard cosmological model. 
The primordial density fluctuations predicted by popular 
inflationary universe models
have very simple characteristics; the fluctuations are
nearly scale invariant and the corresponding 
mass variance is progressively
larger at smaller masses \cite{Riotto}. 
Theoretical studies and numerical simulations of early structure 
formation based on such models
suggest that the first cosmic structures form as early
as when the universe is one hundred million 
years old\cite{Miralda03, Gao07}.
  
Dense, cold clouds of self-gravitating
molecular gas develop in the inner regions of small dark matter 
halos and contract
into proto-stellar objects with masses of about several 
hundreds of solar-masses.
Figure~\ref{ny_fig1} shows the projected gas distribution 
in a cosmological simulation that includes hydrodynamics and 
primordial gas chemistry\cite{Hirano13}.
Star-forming gas clouds are found at
the knots of filaments, which resemble the large-scale
structure of the universe, although actually much smaller in mass 
and size. 

As soon as the first stars are formed, they emit light
and flood the universe with ultra-violet photons.
While some of the gas clouds actually bear stars, other clumps, the
so-called minihalos, remain as neutral gas clouds which might
be significant sinks of photons via recombination processes 
later during the epoch of reionization. \cite{Ahn} 
It is generally thought that cosmic reionization is likely initiated 
by the first generation of stars but that the major role is taken over
by larger and more luminous objects. Thus
the emergence of the first galaxies is a critical event
in the early cosmic history. 

The observational frontier extends beyond $z=6$, reaching 
recently to $z=10$. Utilizing the unprecedented near-IR sensitivity of the 
Wide Field Camera~3 on board the {\it Hubble Space Telescope},
deep images of the Hubble Ultra Deep Field and other fields opened up
a fantastic view into the high-redshift Universe.
The galaxy luminosity function at $z > 6$ has been
derived from the combined observations by HST
and by large ground-based telescopes.\cite{Bouwens, Ellis}
Interestingly, the observed high-redshift galaxies {\it cannot} be 
the major source of reionization.\cite{Robertson}
 This can be easily seen by integrating
the luminosity function down to the faint limit detected\cite{Finkelstein}.
It is thus suggested that very faint (proto-)galaxies 
are needed to ionize the IGM perhaps at $z\sim 10$. 
Other faint sources such as small quasars
may also be worth being considered, 
as we discuss in Section 4.

In the standard $\Lambda$ Cold Dark Matter model, where structure grows
hierarchically,  
the first stars are formed before bigger and more 
luminous galaxies emerge.
Feedback effects from the stars are thus expected to 
play a vital role in setting the scene, {\it i.e., the initial conditions}, 
for first galaxy formation\cite{BrommYoshida}.

\begin{figure}
\begin{center}
\scalebox{0.2}{\includegraphics{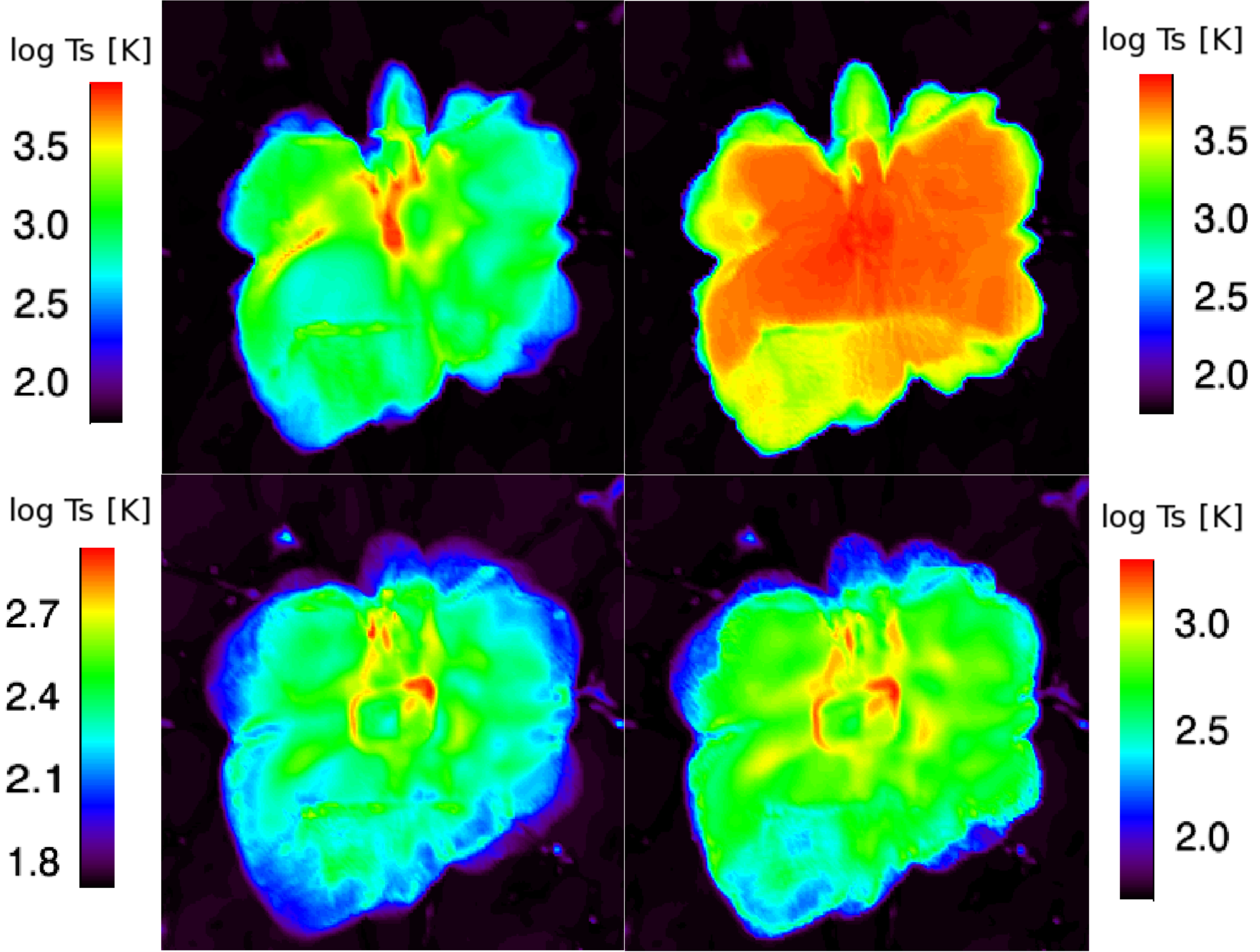}}
\caption{The first \HII region around a massive Population III star.
The plotted region is a cube of 3 kilo-parsecs on a side.
The color scale shows the spin temperature of \cm emission. From Ref.~\citen{tokutani}.
\label{ny_fig2} }
\end{center}
\end{figure}

\section{The first light and \HII regions}
The birth of the first generation of stars has important
implications for the thermal state and chemical properties of the
IGM in the early universe.
As soon as the first stars are formed, they emit a copious amount 
of UV photons and then generate \HII regions.
The formation of early \HII regions were studied by a few groups using
radiation hydrodynamics simulations\cite{Kitayama04, Whalen04}.
It is expected that there are numerous early relic \HII regions formed by
the first stars at $z = 15 - 30$.
Although individual \HII regions are too small and too faint to 
be observed in any wavelength,
they may collectively imprint distinguishable fluctuations 
in the rest-frame \cm.
Figure 2 shows the brightness temperature of an early relic \HII region
in a cosmological simulation\cite{tokutani}. 
The relic \HII region has a large \cm spin temperature 
and thus is bright in radio, yielding a brightness 
temperature of $\sim 1$ mK. However, its physical size of a few kilo-parsecs
hampers direct observations for it to be an individual point source.
Nevertheless, clustering of such \HII regions will
leave detectable imprints in \cm emission\cite{Greif21cm}.

Large-scale \HII regions around galaxies and perhaps early galaxy groups,
extending over tens of mega-parsecs, are probably dominant in volume
at lower redshift of $6<z<10$ where the  ongoing observation
by LOFAR is aimed at. \cite{zaroubi}
Numerical simulations show complex topological features 
of ionized and neutral regions
in a large cosmological volume.
Ultimately, the overall morphology of the \HII bubbles will provide
invaluable information on the sources of reionization and on how the process
occurred in the first one billion years.\cite{Miralda2000}

 Early \HII regions generate secondary CMB anisotropies via 
the kinetic Sunyaev-Zeldovich effect. \cite{hyunbae} 
Figure 3 shows the large-scale
ionization structure and the generated CMB fluctuations calculated
from state-of-the-art $\Lambda$CDM simulations with radiative
transfer. Note that the simulation
covers a volume of more than 100 comoving Mpc on a side. 
Large \HII bubbles are generated not by a single luminous galaxy
but by a group of at least tens of star-forming galaxies.
Highly inhomogeneous distribution of the \HII regions boosts 
the fluctuations of the CMB at small angular scales.
The amplitude and the shape of the angular power-spectrum 
can be used to infer the duration of reionization and the 
overall inhomogeneity of ionized regions as can be seen in Figure 3.

 Recent observations by the South Pole Telescope (SPT) 
collaboration \cite{zahn_etal} have placed 
an upper limit on the CMB temperature fluctuations 
from the kinematic SZ effect 
at $l$ = 3000 to be 
$D_{\rm patchy,3000} < 4.9 \mu$K$^2$ 
at the 95\% confidence level when the degree of angular correlation between 
the thermal Sunyaev-Zeldovich and the cosmic infrared background 
is allowed to vary. The SPT result suggests that reionization 
ended at $z > 5.8$ at 95\% confidence (accounting for the tSZ-CIB correlation), 
in good agreement with other observations. 
We will discuss the current constraints
and the prospects for observations of CMB temperature fluctuations
in more substantial detail later in Section 7.

\begin{figure}
\begin{center}
\scalebox{0.35}{\includegraphics{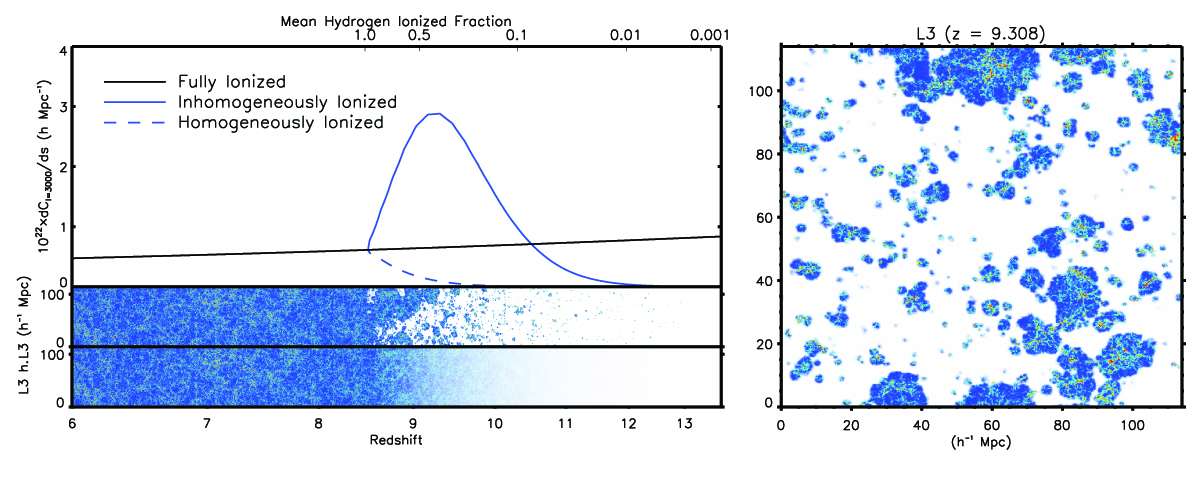}}
\caption{Simulations of the kinematic SZ effect caused by early \HII regions. 
The left panel shows the contribution to the angular power
spectrum from a given redshift. The right panel shows a snapshot
at $z=9.3$ for a volume of 120 Mpc on a side. From Ref.~\citen{hyunbae}. 
\label{ny_fig3} }
\end{center}
\end{figure}

\section{Imprints of dark matter}
There could be sources of reionization other than stars and galaxies,
including somewhat exotic possibilities.  Partial ionization of the
IGM can be caused by X-rays and gamma rays from particle annihilation, and up-scattered CMB photons from inverse Compton scattering \cite{Ricotti, Oh01, arvi1, arvi2, arvi3}.  Chen \&
Miralda-Escud$\acute{\rm e}$ (2008) argue that X-rays from the first
stars heat the surrounding gas and couple the \cm spin temperature to its
kinetic temperature, generating a large Lyman-$\alpha$ absorption
sphere.  X-rays from early mini-quasars could also raise the gas kinetic
temperature and enhance \cm signals.\cite{Kuhlen} Regardless of the
nature of the sources, reionization by X-rays lead to more diffuse
distribution of ionized gases. The particle nature of dark matter may have observable effects on the mass and luminosity of the earliest stars \cite{ds1,ds2,ds3,ds4}.

 High-energy particles and photons that are produced by
decay or annihilation of dark matter (DM) can also ionize the IGM partially.
It is unlikely that significant energy release from DM occurred
early on, because even a small deviation from the well-established
thermal and ionization history of the universe already places
rather tight constraints \cite{galli_cmb, arvi_cmb}. However, if energy release occurred late,
characteristic signals may be imprinted in the CMB
and in the \cm signal.
Recent studies suggest that future radio telescopes can
indeed detect signals from DM annihilation. \cite{Valdes,Finkbeiner}
We will devote more detailed discussion later in Section 7.

 The nature of particle dark matter can affect the early evolution
of the IGM in an indirect but interesting way. 
A combined analysis of high-redshift galaxy number counts, 
other star formation
indicators such as supernovae rate, and the epoch of
reionization can be used to infer the overall growth of sub-galactic
structure in the Dark Ages \cite{bullock}. 
 
It is known that models with warm dark matter, in which
dark matter particles possess substantial thermal motions,
predict less abundant small-scale structures. 
The fact that the universe was reionized early on strongly
suggests that structure formation and the associated star formation
must have occurred similarly early.
Accurate measurement of the Thomson optical depth of the CMB
and also of the visibility function (the derivative of the optical
depth with respect to redshift)
can give constraints on the nature of dark matter, if the derived
optical depth is sufficiently large\cite{YoshidaWarm}. 

\section{Infrared background}
The extragalactic infrared background (IRB) is largely contributed 
by accumulated light emitted from galaxies and quasars. 
The local source, most significantly the
Zodiacal light, and the stellar emission from low-redshift
galaxies are the two dominant sources, but 
the remaining IR flux may be either from high-redshift galaxies
or from low surface brightness galaxies in the local universe.
An interesting possibility is that, 
if Pop~III stars were formed at $z=10-20$, UV photons emitted
from them, redshifted to 1-5 $\mu$m in the present-day universe, 
are expected to contribute to the IRB. In principle, the IRB
can be used to constrain the star formation activity in the
early universe, which should be consistent with
what other probes of cosmic reionization suggest.
Interestingly, a recent cross-correlation analysis of IRB and X-ray 
indicate that AGNs or some IR sources associated with them 
contributes appreciably 
to about 10 percent of the IRB.\cite{Cappe}

 It has been long speculated from an apparent bump
in the IRB at 1-2 micron\cite{IRTS} that a significant amount
of hot Population III stars were formed at $z\sim 10$
(see Figure 4).
The bump, if real, can be explained largely by photons 
near Lyman-$\alpha$ wavelength redshifted from $z=10$ to $z=0$. 
Such an intense formation of massive Population III stars
at $z\sim 10$, when substantial metal enrichment must
have already occurred, is not expected in popular
models of first star formation. 
However, it remains still controversial 
whether or not an early generation of stars and galaxies
contribute to the IRB.\cite{Fernandez}
The overall amplitude of the IRB, estimated from the
spectra of distant TeV blazers for example, still allows
small contribution from unknown sources \cite{TeV}, 
and the IRB fluctuations measured by Spitzer and Akari
both suggest that the fluctuation power spectra can be
explained by clustered sources at high redshifts.\cite{Kashlinsky, Akari}
The low-level shot noise features and the shape of the 
power spectra at sub-degree scales 
may be reconciled by intra-halo stars
around galaxies at $z=1-4$. \cite{CoorayNature}
A concerted use of multi-wavelengths observations 
will be needed to distinguish various models 
and pin down the sources of reionization.
 
\begin{figure}
\begin{center}
\scalebox{0.4}{\includegraphics{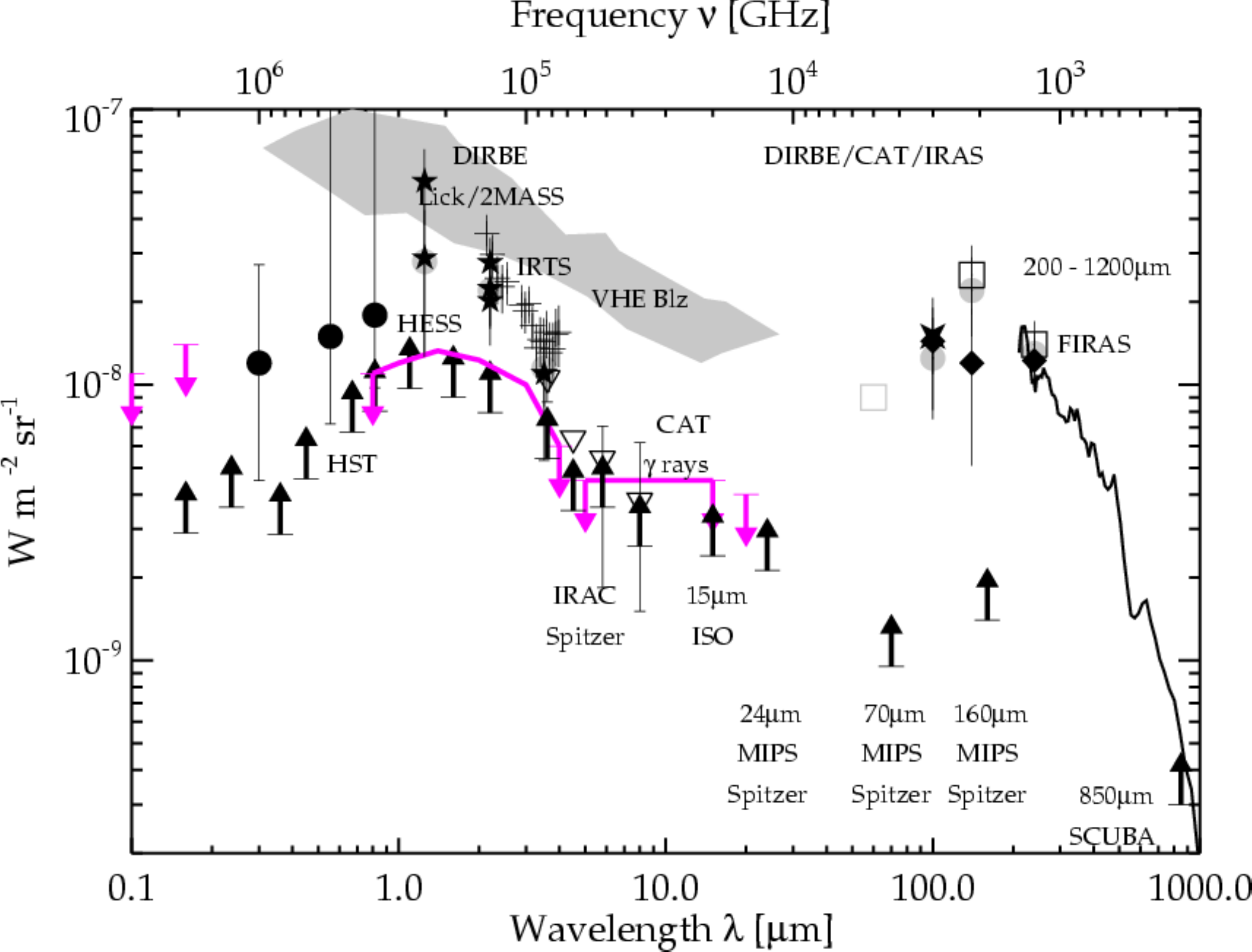}}
\caption{The observed spectrum of the cosmic infrared background.
From Ref. \citen{Dole06}.
\label{ny_fig3} }
\end{center}
\end{figure}


\section{Probes of reionization}

From observations of the spectra of distant quasars, it is known that the Universe is highly ionized today \cite{fan1, fan2, becker, pentericci}. Evidence for reionization at the $5.5\sigma$ level was obtained by the Wilkinson Microwave Anisotropy Probe (WMAP) measurement of the CMB EE polarization power spectrum \cite{wmap}. 

Reionization began at a redshift $z \sim 20 - 30$ when the first stars were formed. Later, Population II stars, star forming galaxies, and active galactic nuclei completed the process \cite{tumlinson_shull, loeb_barkana_2001, barkana_loeb_2001, wyithe, ciardi, sokasian}. The precise details of the reionization process are not known, and must be inferred from observations. Let us now discuss two promising probes of reionization - the \cm spin flip of neutral hydrogen, and the cosmic microwave background.

\subsection{ Probing the dark ages through \cm observations}

The nature of the earliest stars is a fascinating topic, but one which is very difficult to study due to the lack of observations. Emission and absorption due to the \cm spin flip transition of neutral hydrogen have emerged as useful techniques to probe the epoch of primordial star formation \cite{madau, loeb_zald, cooray, bharadwaj_ali, carilli, furl_briggs,furl_briggs,pritchard_loeb2010, pritchard_loeb2012,liu_paper}.  Before the formation of the first luminous objects, the spin temperature of neutral hydrogen is typically close to the CMB temperature at a redshift $z \sim 20$ because the gas is not dense enough to collisionally couple the spin temperature to its kinetic temperature \cite{loeb_zald}. The formation of the first stars however, results in the production of Lyman-$\alpha$ photons which can couple the spin and kinetic temperatures of neutral hydrogen through the Wouthuysen-Field mechanism \cite{wou,fie}. The kinetic temperature of the gas $T_{\rm k} \propto (1+z)^2$ is typically lower than the CMB temperature $T_\gamma$ (which scales as (1+$z$)) at $z\sim20$. 

The Wouthuysen-Field mechanism sets $T_{\rm s} = T_{\rm k}$, so the turn-on of the first stars produces a significant decrease in the \cm brightness temperature ($T_{\rm b}$) around the redshift of first star formation, with $T_{\rm b} \propto (T_{\rm s} - T_\gamma )/T_{\rm s} \sim - T_\gamma /T_{\rm k}$ . We thus expect a significant decrement in the \cm brightness temperature $T_{\rm b}$ around the redshift of first star formation, since $T_{\rm b} \propto (T_{\rm s} - T_\gamma) / T_{\rm s}$. Heating of the gas by ionizing radiation rapidly sets $T_{\rm k} > T_\gamma$, with $T_{\rm b}$ entering the saturation regime before decreasing to zero as the Universe reionizes. The magnitude of the decrement in $T_{\rm b}$ as well as the width provide valuable information on the properties of the first stars and X-ray sources. The spectral structure allows us to distinguish the signal from the much larger background which is spectrally smooth.

The \cm brightness temperature relative to the CMB (also called differential 
brightness temperature) is given by (see, for example \cite{Furlanetto}):
\begin{equation}
T_{\rm b} \approx 27 \, {\rm mK} \; x_{\rm HI} \; \sqrt{ \frac{1+z}{10} } \, \left( 1 - \frac{T_\gamma}{T_{\rm s}} \right ),
\label{21cmtemp}
\end{equation}
where $x_{\rm HI}$ is the neutral hydrogen fraction, and $T_{\rm s}$ is the spin 
temperature of neutral hydrogen given by:
\begin{equation}
T^{-1}_{\rm s} \approx \frac{ T^{-1}_\gamma + \left( x_{\rm c} + x_\alpha \right ) T^{-1}_{\rm k} }{1 + x_{\rm c} + x_\alpha}.
\label{xi}
\end{equation}
\begin{figure}[!t]
\begin{center}
\scalebox{0.3}{\includegraphics{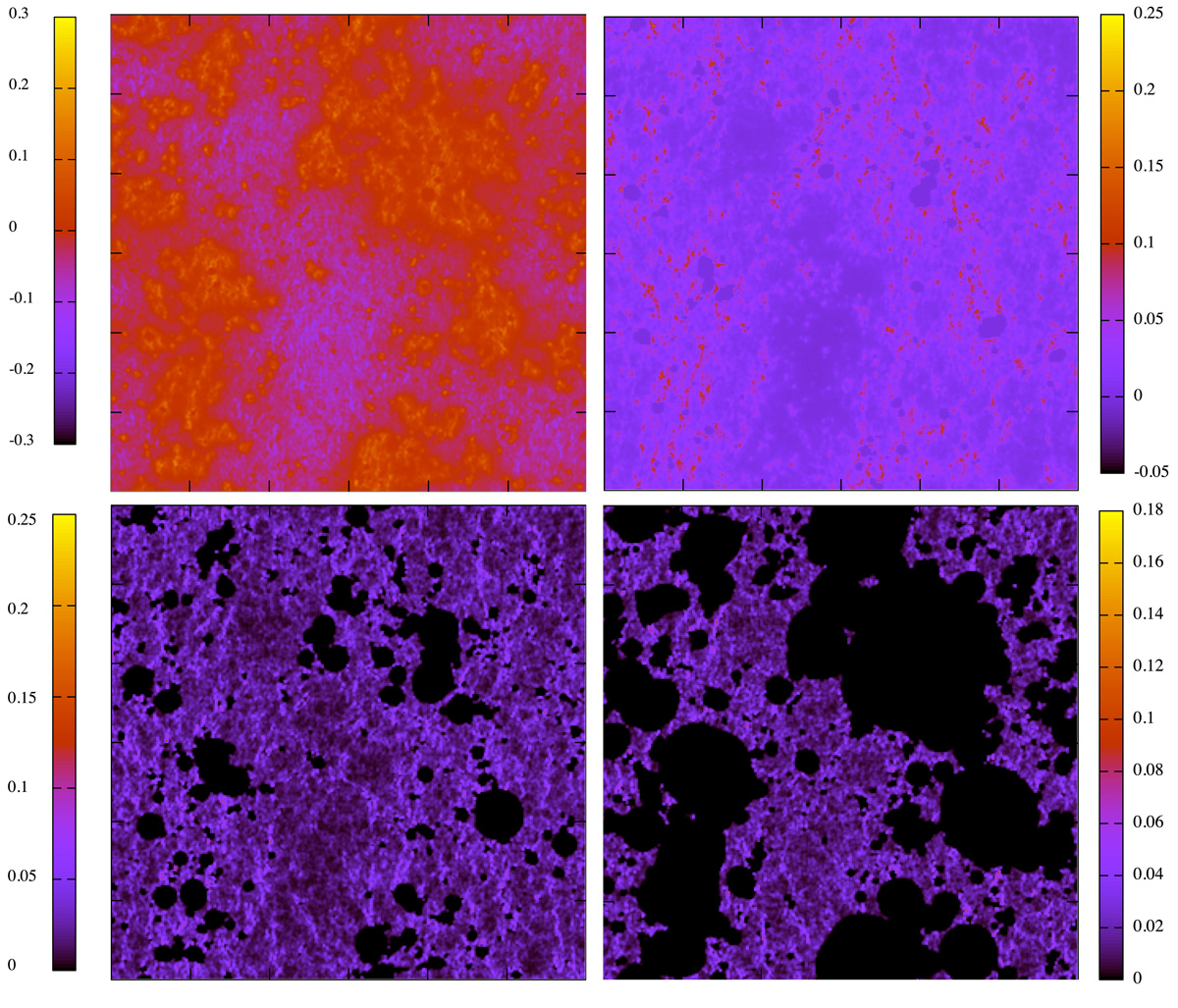}}
\end{center}
\caption{The large-scale \cm fluctuations calculated by SIMFAST.
Each panel is 300 Mpc $\times$ 300 Mpc, and shows the brightness temperature in Kelvin 
at $z = 14, 12, 10, 8$ (from top left to bottom right).
\label{arvi_fig0} }
\end{figure}

\begin{figure}[!t]
\scalebox{0.5}{\includegraphics{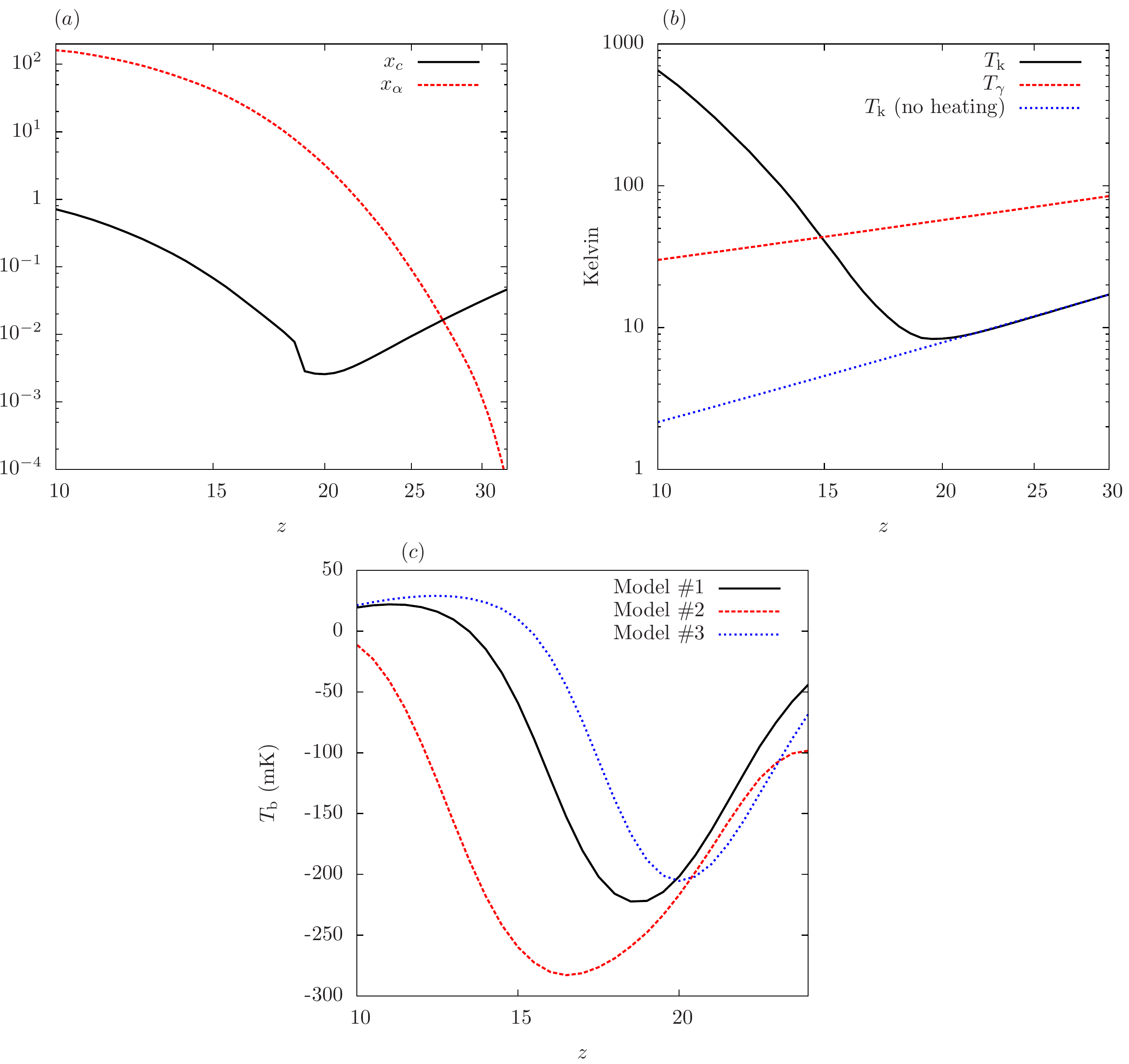}}
\caption{ Panel (a) shows the collisional coupling coefficient compared to the Lyman-$\alpha$ coupling coefficient. At high redshifts, collisional coupling provides  the dominant contribution, but after first star formation, $x_\alpha \gg x_{\rm c}$. (b) shows the gas kinetic temperature (solid, black) and the CMB temperature (dashed, red). The dotted (blue) line shows the gas temperature in the absence of heating. (c) shows the \cm brightness temperature (relative to the CMB), for three different models that differ in their star formation rate, and  X-ray flux.
\label{arvi_fig1} }
\end{figure}

$x_{\rm c}$ is called the collisional coupling coefficient, while $x_\alpha$ is  the Lyman-$\alpha$ coupling due to the Wouthuysen-Field mechanism.  Collisional coupling is important when the gas is dense or hot, i.e.  at high redshifts.  Figure \ref{arvi_fig1}(a) shows the collisional coupling and Lyman-$\alpha$ coupling ($x_\alpha$) as functions of redshift (for a particular star formation model). Once star formation begins at a redshift $z \lesssim 30$, the Lyman-$\alpha$ coupling provides the dominant contribution. Figure \ref{arvi_fig1}(b) shows the gas kinetic temperature for a specific star formation model, as well as the CMB temperature. At very high redshifts ($z \gtrsim 300$), the gas temperature closely follows the CMB temperature due to Compton scattering with residual electrons. At lower redshifts, the gas is not sufficiently dense for Compton scattering to be efficient. The kinetic temperature then falls off $\propto (1+z)^2$ until star formation begins. Three dimensional maps of the brightness temperature of neutral Hydrogen can be used to infer the reionization history of the Universe. Fig. \ref{arvi_fig0} shows the simulated brightness temperature (for a particular star formation model) in 300 Mpc $\times$ 300 Mpc boxes at redshifts $z$ = 14, 12, 10, and 8, obtained using the {\scriptsize SIMFAST} code \cite{simfast}.

The intensity of radiation in the Lyman-$\alpha$ wavelength at any given redshift $z$ is due to radiation emitted between the Lyman-$\alpha$ wavelength and the Lyman limit.  Thus radiation emitted at wavelengths shorter than Lyman-$\alpha$, at a redshift $z' < z_{\rm max}(n)$ will redshift until the photons reach the Lyman-$\alpha$ wavelength at a redshift $z < z'$. Thus photons with wavelengths between Lyman-$\alpha$ and Lyman-$\beta$ are visible up to a maximum redshift $z_{\rm max}(2)$, where $1 + z_{\rm max}(2) = (\lambda_\alpha/\lambda_\beta)(1+z)$ \cite{barkana_loeb2005}. The Lyman-$\alpha$ coupling coefficient $x_\alpha$ is proportional to the Lyman-$\alpha$ photon intensity \cite{barkana_loeb2005}:
\begin{equation}
J_\alpha = \frac{(1+z)^2}{4\pi} \sum_{n=2}^{n_{\rm max}} \int_z^{z_{\rm max}(n)} \; \frac{c dz'}{H(z')} \epsilon(\nu'_n,z'),
\end{equation}
where $\epsilon$ is the number of photons emitted per comoving volume, per unit time, per frequency, and depends on the nature of the ionizing sources. 

Once the Lyman-$\alpha$ coupling $x_\alpha \gg 1$, the spin temperature $T_{\rm s}$ is set to the kinetic temperature $T_{\rm k}$ of the gas. The redshift at which $x_\alpha > 1$ is however, sensitive to the nature of the ionizing sources. The initial mass function (IMF) of Pop. III plays an important role in determining the number of Lyman-$\alpha$ and ionizing photons.  Authors \cite{muratov_etal} find that a 170 $M_\odot$ star emits about 34,500 ionizing photons per baryon over its lifetime, compared to $\approx$ 6600 per baryon per lifetime for a Pop. II star. Authors \cite{bromm2001} find that a heavy IMF with $M > 300 M_\odot$ produces 16 times as many ionizing photons compared to a Salpeter IMF. Thus, one may hope to place constraints on the nature of Pop. III stars by measuring the brightness temperature of neutral Hydrogen \cm radiation, although it will be challenging to break the degeneracy between the primordial star IMF and the star formation rate \cite{simfast}

Accretion of gas onto black holes produced by the first stars will generate highly energetic X-rays which heat the gas to temperatures above the CMB. The temperature evolution of the gas in the presence of X-ray heating is given by:
\begin{equation}
-(1+z)H(z)\frac{dT_{\rm k}}{dz} = -2T(z)H(z) + \frac{2\eta_{\rm heat}(z)}{3 k_{\rm b}} \xi(z),
\end{equation}
in the limit of ionized fraction $x_{\rm ion} \ll 1$. $\xi(z)$ is the energy absorbed per atom per unit time at redshift $z$. $\eta_{\rm heat}$ is the fraction of the absorbed energy that goes into heating. Detailed computations \cite{furl_stoever, shull_steen, kk, valdes}  show that $\eta_{\rm heat}$ is a function of photon energy, as well as the ionized fraction. For highly neutral gas $x_{\rm ion} \approx 10^{-4}$, we have $\eta_{\rm heat} \lesssim 0.2$ for photon energies $E_\gamma > 100$ eV. For slightly ionized gas with $x_{\rm ion} \sim 0.01$, $\eta_{\rm heat} \sim 0.4$ for $E_\gamma > 100$ eV, with lower energy photons contributing more to heating \cite{furl_stoever}.

 The ratio of temperatures in the absence of any heating is approximately given by $T_\gamma / T_{\rm k} \approx 7.66 \; \left[ 20 / (1+z) \right ]$. From Eq. \ref{21cmtemp}, it is easy to see that the minimum brightness temperature (relative to the CMB) is  $ \approx -300 \, \sqrt{20/(1+z)}$ mK, when the gas is not heated by X-rays, and when the spin and kinetic temperatures are well coupled by Lyman-$\alpha$ photons. Figure \ref{arvi_fig1}(c) shows the brightness temperature of neutral hydrogen relative to the CMB, obtained using the {\scriptsize SIMFAST} code \cite{simfast1,simfast}, for three different models that differ in their star formation efficiency and X-ray heating flux. The location of the trough in $T_{\rm b}$, as well as its width are determined by the physics of primordial star formation, i.e. the Lyman-$\alpha$, and X-ray flux, which in turn may be related to the star formation efficiency, and X-ray heating rate. It is clear that a precise measurement of the \cm temperature will provide important information regarding the formation of the first stars.

\begin{figure}[t]
\scalebox{0.5}{\includegraphics{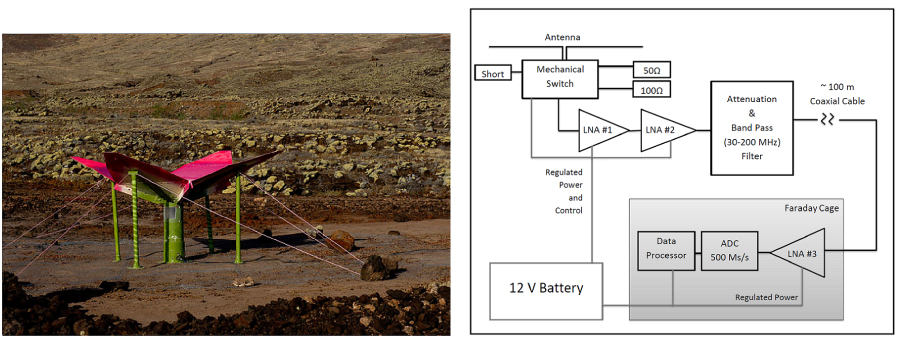}}
\caption{ The SCI-HI experiment showing the antenna on-site at Isla Guadalupe, and the system block diagram. Figure from Ref.\citen{hibiscus}.
\label{arvi_fig2} }
\end{figure}

\subsection{Measuring the global \cm brightness temperature}

Experiments studying the universe through the \cm transition include the Precision Array for Probing the Epoch of Reionization (PAPER) \cite{pober,parsons}, the Giant Metrewave Radio Telescope - Epoch of Reionization (GMRT-EoR) \cite{paciga}, the Low Frequency Array (LOFAR) \cite{vanh}, the Murchison Widefield Array (MWA) \cite{bernardi}, the Hydrogen Epoch of Reionization Array (HERA) \cite{hera}, the Square Kilometer Array (SKA) \cite{ska}, the Experiment to Detect the Global EoR Step (EDGES) \cite{edges},  the  Large Aperture Experiment to detect the Dark Ages (LEDA) \cite{leda1}, the Sonda Cosmol\'{o}gica de las Islas para la Detecci\'{o}n de Hidr\'{o}geno Neutro (SCI-HI) \cite{hibiscus}, and the Dark Ages Radio Explorer \cite{dare}. \\ 

\emph{The SCI-HI \cm all-sky spectrum experiment:} \\

The Sonda Cosmol\'{o}gica de las Islas para la Detecci\'{o}n de Hidr\'{o}geno Neutro (SCI-HI) experiment consists of a single broadband sub-wavelength size antenna and a sampling system for real time data processing and recording \cite{hibiscus}.  Preliminary observations were completed in June 2013 at Isla Guadalupe, a Mexican biosphere reserve with minimal infrastructure, located $\sim$ 260 km from the Pacific coast.

Figure \ref{arvi_fig2} shows the antenna on site, as well as a basic block diagram of the instrument. The signal from the antenna passes through a series of electronic stages, including amplifiers and filters to remove radio frequency interference (RFI) below 30 MHz and aliasing of signals above 200 MHz. The system is placed inside a Faraday cage $\sim$50 meters from the antenna. The data sampling/processing duty cycle is $\sim$10\%, so 1 day of observation yields about 2 hours of effective integration time. 

\begin{figure}[t]
\scalebox{1.15}{\includegraphics{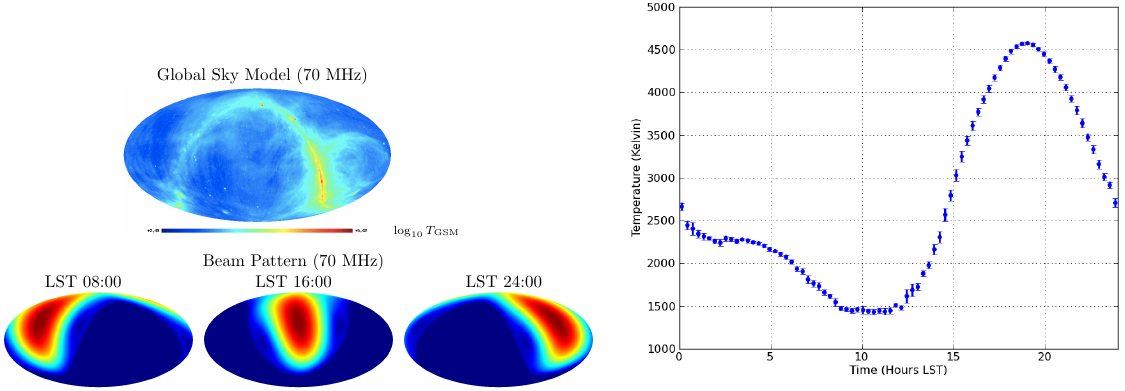}}
\caption{ Sky temperature and antenna beam pattern in (RA,DEC) coordinates. The top row shows the sky temperature (logarithmic units) at 70 MHz, from Ref. \citen{angelica}. Also shown is the simulated antenna beam pattern at 70 MHz at different LST, plotted for the latitude of Isla Guadalupe. Shown on the right is the diurnal variation of a single 2 MHz wide bin centered at 70 MHz, for 9 days of observation, binned in 18 minute intervals. Figure from Ref. \citen{hibiscus}.
\label{arvi_fig3} }
\end{figure}

Calibration of data is performed by comparing the measured brightness temperature to the Global Sky Model of the Galaxy.  Figure \ref{arvi_fig3} shows the Global Sky Model of the Galaxy, at 70 MHz, from Ref.\cite{angelica}. Also shown is the simulated beam pattern of the antenna at Local Sidereal Time (LST) 08:00, 16:00, and 24:00, plotted for the latitude of Guadalupe.  The antenna beam is fairly broad $\sim 55^\circ$ at 70 MHz, and thus, this is low angular resolution experiment. The antenna beam is averaged over the Galaxy. We therefore expect a large sky brightness when the Galaxy is overhead, and a minimum when the Galaxy is aligned with the horizon. This is precisely what we see in the plot to the right. The diurnal variation of the Galactic temperature may be used to subtract the large Galactic foreground, to recover the cosmological \cm signal which does not vary with time. The calibrated spectrum is fit to the Galactic sky-averaged brightness temperature ($T_{\rm GM}$):
\begin{equation}
\log_{10} T_{\rm GM} (\nu) = \sum_{k=0}^{n} a_k \, \left [ \log_{10} \left( \frac{\nu}{70 \, {\rm MHz}} \right ) \right ]^k
\end{equation}
Using the calculated $a_k$ for each day of data, the residuals $\Delta T(\nu) = \langle T_{\rm meas} \rangle_{\rm DAY}(\nu) - T_{\rm GM} (\nu)$ are calculated. These $\Delta T(\nu)$ values are our estimate of the \cm all-sky brightness temperature spectrum after removal of Galactic emission.  An $n=2$ fit captures the band average expected foreground brightness temperature ($a_0$), a power law spectral shape ($a_1$), and a self-absorption correction term ($a_2$). Adding additional $a_k$ terms is found to have minimal impact on the overall residual levels.

Residuals obtained after subtraction of the large Galactic foreground are $\lesssim$ 20 Kelvin from 4.4 hours of integration, in the range 60-88 MHz ($15<z<23$). Given that the mean foreground is between 2000 - 5000 Kelvin, the residuals obtained by SCI-HI are $< 1\%$ of the foreground signal. Nevertheless the residuals obtained are nearly two orders of magnitude larger than the cosmological \cm brightness temperature. Improvements in system design are currently underway. Data collection is expected to resume in the Summer of 2014, at Isla Socorro, or Isla Clari\'{o}n which are exceptionally radio  quiet sites. \\

\emph{The Dark Ages Radio Explorer:} \\

Even the most remote regions on Earth suffer from some man-made radio frequency interference, as well as ionospheric effects. The Dark Ages Radio Explorer (DARE) \cite{dare} is a space based cosmology mission that aims to detect the redshifted \cm brightness temperature in the frequency range 40-120 MHz ($11<z<34$).  
DARE will orbit the moon for a mission lifetime of 3 years, and will collect data above the lunar farside, free from radio frequency interference, ionospheric effects, and heliospheric emissions.  Thus, DARE is expected to shine light on primordial star formation: in particular DARE expects to place useful bounds on the epoch of first star formation, the formation of the first accreting black holes, and the start of the reionization epoch.


The DARE radiometer consists of a dual-polarized antenna with a compact, integrated, front-end electronics package, a single-band, dual-channel receiver, and a digital spectrometer. The antenna consists of a pair of bi-conical dipoles, made unidirectional by a set of deployable radials attached to the spacecraft bus that act as an effective ground plane. The receiver provides amplification of the antenna signal to a level sufficient for further processing by the digital spectrometer, and incorporates a load switching scheme to assist calibration. There are two receivers to accommodate the antennas, i.e.  one receiver per antenna polarization.  Noise diodes provide a reliable additive noise temperature during operation. The noise diodes will be used to monitor spectral response and linearity of downstream components. The instrument design is coupled to a multi-tiered calibration strategy to obtain the RF spectra from which the \cm signal can be extracted. The calibration strategy relies on the fact that a high degree of absolute calibration is not required, but focuses on the relative variations between spectral channels, which are much easier to control.

The antenna power pattern covers approximately $1/8$ of the sky depending on frequency, and the data set will consist of spectra from 8 independent regions on the sky. DARE uses four free parameters to fit the foreground, i.e. $\log T_{\rm FG} = \log T_0 + a_1 \log \nu +  a_2 \left ( \log \nu  \right )^2 + a_3 \left ( \log \nu  \right )^3$. The parameter values are fit separately, to each sky region. Thus, there are 32 foreground parameters in total. It is possible to separate the signal from the large foregrounds because the foregrounds are spectrally smooth, while the \cm brightness temperature has spectral structure. Also the foregrounds are spatially varying, while the \cm temperature is spatially smooth. With three years of observation, the DARE mission is expected to obtain 3000 hours of integration. The experiment is expected to constrain the epoch of first star formation to $\sim$ 9\% accuracy, the start of X-ray heating by accreting black holes to $\sim$ 1.4\% accuracy, and the start of reionization to $\sim$ 0.4\% accuracy.

Both SCI-HI and DARE expect to measure the \emph{global} \cm signal. The \emph{power spectrum} of \cm fluctuations is also a useful tool to study reionization. There exist fluctuations in the \cm brightness temperature due to fluctuations in the matter density, neutral fraction, and temperature. Fluctuations in the \cm brightness temperature are expected to be large near the edges of HI regions, and therefore may be more easily separated from the large Galactic foreground compared to the global signal, particularly near the end of reionization. The fluctuation in the brightness temperature $\delta_{21} = \delta T_{\rm b} / T_{\rm b}$ is caused by fluctuations in the density field $\delta$, the neutral fraction $\delta_{\rm HI}$, the Lyman-$\alpha$ flux $\delta_\alpha$, the radial velocity gradient $\delta_{d_rv_r}$and the temperature $\delta_{\rm T}$ \cite{barkana_loeb_2005_1}:
\begin{eqnarray}
\delta_{\rm 21} &=& \left[ 1 + \frac{x_{\rm c}}{x_{\rm tot} (1 + x_{\rm tot})} \right ] \delta + \frac{x_\alpha}{x_{\rm tot} (1 + x_{\rm tot})} \delta_\alpha + \delta_{\rm HI}  \nonumber \\
&-& \delta_{d_rv_r} + \delta_{\rm T} \left( \frac{T_{\rm k}}{T_{\rm k} - T_\gamma} + \frac{x_\alpha}{x_{\rm tot} (1 + x_{\rm tot})} \frac{d \log \kappa_{10}}{d \log T_{\rm k}} \right ),
\label{long_eqn}
\end{eqnarray}
where $x_{\rm tot} = x_{\rm c} + x_\alpha$, and $\kappa_{10}(T_{\rm k})$ is the collisional spin de-excitation rate coefficient. The fluctuations in Fourier space may be expressed in the form\cite{barkana_loeb_2005_1}:
\begin{equation}
\tilde \delta_{T_{\rm b}}(\vec k) = \mu^2 \tilde\delta(\vec k) + \beta \tilde\delta(\vec k) + \tilde\delta_{\rm rad}(\vec k),
\end{equation}
where $\mu$ is the cosine of the angle between the wave vector $\vec k$ and the line of sight, while $\beta$ is obtained by collecting together the terms in Eq. \ref{long_eqn}.
 Early results on reionization have been obtained by the PAPER experiment \cite{parsons} using the power spectrum of \cm observations. The PAPER experiment consists of 32 antennas, and collects data in South Africa. With an exposure of 55 days, PAPER obtained an upper limit on the \cm power spectrum of $\left ( 52 {\rm mK} \right )^2$ for $k$ = 0.11 $h$/Mpc at $z$=7.7 at the $2\sigma$ level.


\section{Probing reionization with the cosmic microwave background}

So far, we have discussed how the highly redshifted \cm radiation from neutral Hydrogen can probe early reionization.  The cosmic microwave background is also an excellent probe of reionization. This is because free electrons scatter microwave photons, modifying the spectrum of anisotropies, and generating new, secondary anisotropies.  

\begin{figure}[t]
\begin{center}
\scalebox{1.}{\includegraphics{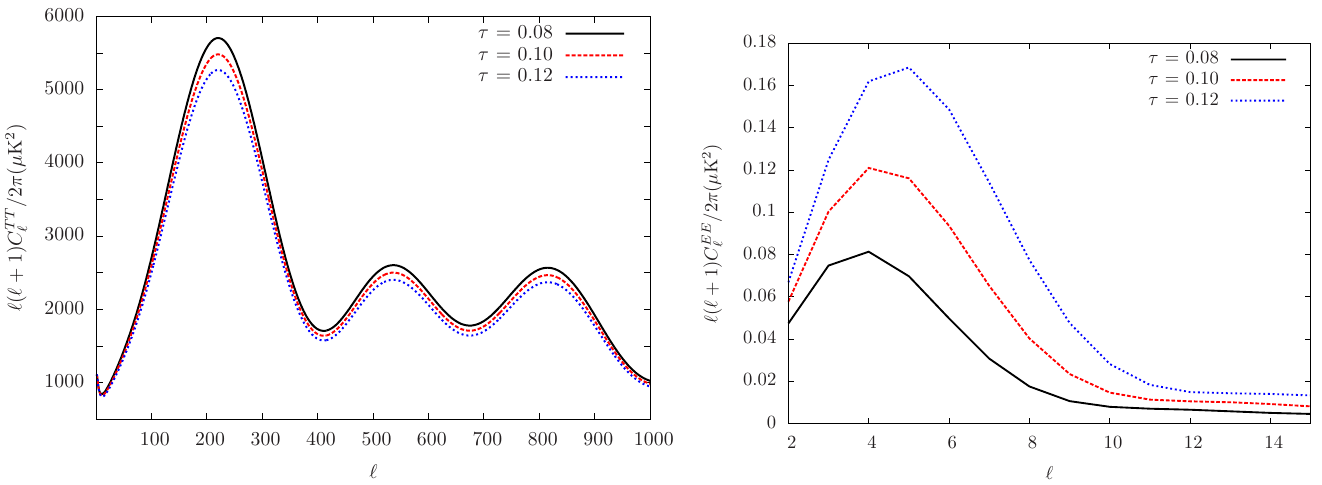}}
\end{center}
\caption{ CMB power spectra - $TT$ and $EE$ for different values of $\tau$. Scattering of CMB photons with free electrons results in a damping in the $TT$ power spectrum, and a boost in the large angle $EE$ power spectrum proportional to $\tau^2$.
\label{arvi_fig5} }
\end{figure}
 
Scattering of CMB photons by free electrons leads to a damping in the temperature power spectrum by a factor $\exp\left[ -2\tau \right ]$. Unfortunately, this damping is largely degenerate with the amplitude of the primordial curvature power spectrum ($A_{\rm s}$ or $\Delta^2_{\rm R}$). This degeneracy is broken by polarization. Thomson scattering polarizes the CMB, and hence leads to a \emph{boost} in the large angle $EE$ power spectrum. Figure \ref{arvi_fig5} shows the $TT$ and $EE$ power spectra for three different values of $\tau$. The Planck experiment (with WMAP polarization data included) has obtained a value of optical depth $\tau = 0.089^{+0.012}_{-0.014}$. The mean redshift of reionization is then $z_\ast = 11.1 \pm 1.1$. The \emph{duration} of reionization is however, not well constrained by the polarization power spectrum.

Secondary scattering of CMB photons introduces additional power on small scales. High energy photons emitted from luminous sources ionize the region around the sources, forming bubbles of hot ionized gas. Reionization is then said to be patchy, and the optical depth becomes a function of direction, i.e. $\tau = \tau(\hat n)$, and can be described to lowest order by 2 quantities: the mean over angles $\langle \tau \rangle$, and the variance over angles, or equivalently, the root mean square value $\tau_{\rm RMS}$. In the presence of a patchily reionized Universe, one observes a different CMB temperature when the line of sight passes through an ionized region, compared to a neutral region. An anisotropic optical depth therefore introduces ``patchy screening'' of the CMB, and hence secondary CMB power on small scales.

\begin{figure}[t]
\begin{center}
\scalebox{0.9}{\includegraphics{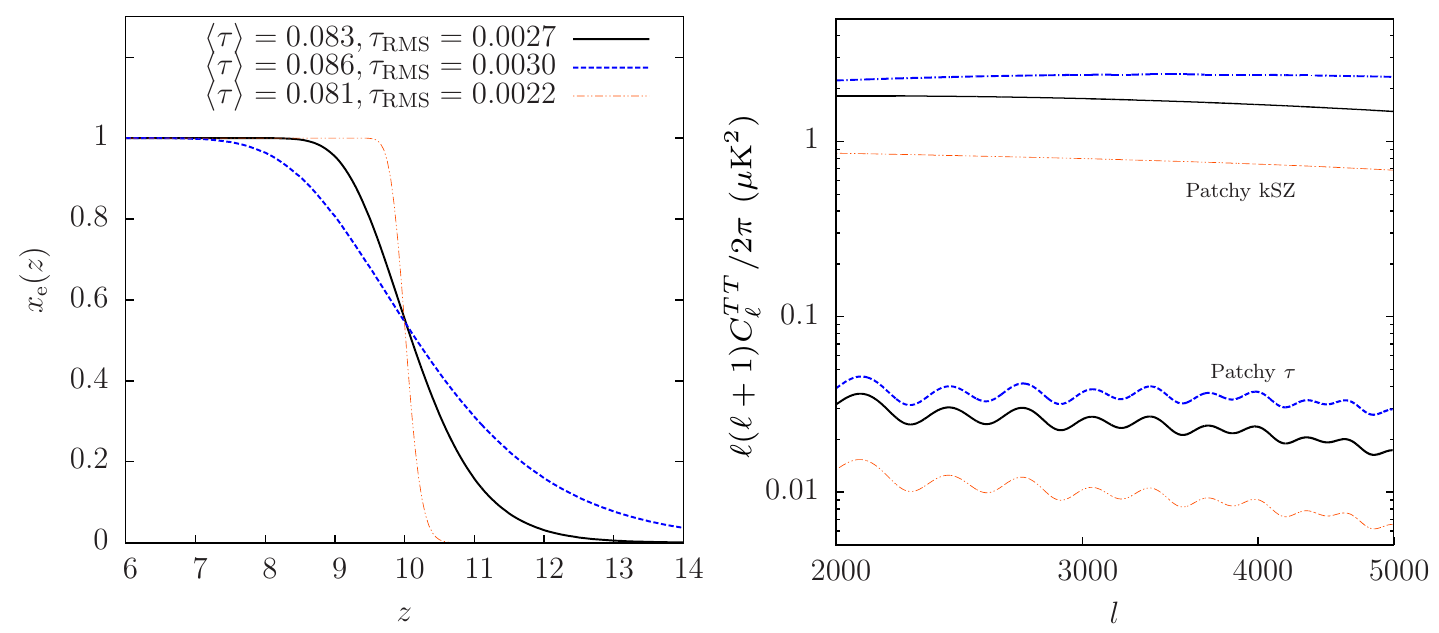}}
\end{center}
\caption{
Shown are three different reionization scenarios that have the same mean redshift of reionization, but different values of $\langle\tau\rangle$ and $\tau_{\rm RMS}$. The plot on the right shows the corresponding kSZ power spectra, and patchy $\tau$ power spectra.  From Ref. \citen{paper2} 
\label{arvi_fig6} }
\end{figure}

CMB photons scattering off moving electrons also introduces power on small scales, the well known Sunyaev-Zeldovich effect \cite{sz1, sz2}. When the electron velocity is due to thermal motion of gas atoms, it is known as the thermal Sunyaev-Zeldovich (tSZ) effect. The main contribution to the tSZ power comes from galaxy clusters. Scattering of CMB photons by free electrons with a bulk velocity (i.e. velocity relative to the Hubble flow) results in the kinetic Sunyaev-Zeldovich (kSZ) power. The fractional temperature change induced by electrons with a bulk motion along the line of sight is \cite{vish1,vish2}:
\beq
\frac{\Delta T}{T} = - \int cdt \left( \hat n \cdot \frac{\vec v}{c} \right ) n_{\rm e} \sigma_{\rm T} e^{-\tau},
\eeq
where $\hat n$ is a unit vector denoting the line of sight, $\vec v$ is the peculiar velocity of the electrons, $n_{\rm e}$ is the number density of free electrons, $\sigma_{\rm T}$ is the Thomson cross section, and $\tau$ is the optical depth. The homogeneous, linear contribution to the kSZ is called the Ostriker-Vishniac effect \cite{vish1,vish2}. In this approximation, the peculiar velocity $\vec v$ may be simply expressed in terms of the matter overdensity. The Ostriker-Vishniac power spectrum may then be analytically computed (see for e.g., \cite{jaffe_kamion} and \cite{kheeganlee}). The total kSZ is almost always larger than the Ostriker-Vishniac power due to non-linearities, and patchiness in the reionization field. The patchy component of the kSZ due to a patchily reionized Universe  is  a good probe of the \emph{duration} of reionization.

Figure \ref{arvi_fig6} (left) shows three different non-instantaneous reionization scenarios, assuming single step reionization (from Ref. \cite{paper2}). These three reionization histories have the same mean reionization redshift, but different mean values $\langle\tau\rangle$, as well as different durations, and hence different values of $\tau_{\rm RMS}$. The short  reionization scenario has $\langle\tau\rangle$ = 0.081, $\tau_{\rm RMS}$ = 0.0022, the fiducial model has $\langle\tau\rangle$ = 0.083, $\tau_{\rm RMS}$ = 0.0027, and the extended reionization model has $\langle\tau\rangle$ = 0.086, $\tau_{\rm RMS}$ = 0.0030.  These values of $\tau_{\rm RMS}$ are representative of what is seen in realistic numerical simulations. The plot on the right shows the corresponding values of patchy kSZ as well as excess power due to patchy $\tau$, for the three reionization models. Using large scale numerical simulations, authors \cite{paper1, paper3} found an approximate scaling relation for the patchy kSZ power:
\begin{equation}
D^{kSZ}_{\ell=3000} \approx 2.02 \, \mu{\rm K}^2 \; \left[ \left( \frac{1+\bar z}{11} \right ) - 0.12 \right ] \left( \frac{\Delta_{\rm z}}{1.05} \right )^{0.47},
\end{equation}
where $D_\ell = l(l+1)C_l/2\pi$, and $\Delta z = \left [ z(x_{\rm e}=25\%) - z(x_{\rm e}=75\%) \right ]$.

\subsection{Detecting patchy reionization through cross correlation of the CMB}

Let us now consider a different technique to probe patchy reionization. Since the damping term $\exp \left[-\tau(\hat n) \right ]$ multiplies the primary CMB temperature $T$, the effect of patchy reionization is largest when $|T|$ is large, i.e. on degree scales. \emph{Patchy reionization therefore transfers CMB power from large scales to small scales}. This results in a non-zero correlation between large and small scales. This ``patchy $\tau$ correlator'' is far more sensitive to patchy reionization than the power spectrum.

To compute the patchy $\tau$ correlator, we begin by filtering the CMB into 2 maps:  (i) A map with information only on large scales, i.e. multipoles $\ell < \ell_{\rm boundary1}$, and (ii) A map with information only on small scales, $\ell > \ell_{\rm boundary2}$.  The 2 maps are then squared:
$f = T^2 (\ell < \ell_{\rm boundary1}), g = T^2(\ell > \ell_{\rm boundary2})$. The patchy $\tau$ correlator is then simply $\langle \delta f \delta g \rangle$, where $\delta f = f - \langle f \rangle$, and $\delta g = g - \langle g \rangle$ are fluctuations in the squared CMB temperature obtained from the filtered maps. The angle brackets denote an average over the map. Let us first examine a simple model wherein we ignore all secondaries besides patchy reionization. Let $\theta_{\rm obs}(\hat n)$ be the observed CMB temperature fluctuation, and let $\theta_{\rm cmb}(\hat n)$ be the primordial CMB fluctuation. $\theta_{\rm cmb}$ consists of large scale and small scale modes, i.e. $\theta_{\rm cmb} = \theta_{\rm L} + \theta_{\rm S}$. The optical depth is $\tau(\hat n) = \langle \tau \rangle + \delta \tau(\hat n)$ .

\begin{figure}[t]
\begin{center}
\scalebox{0.5}{\includegraphics{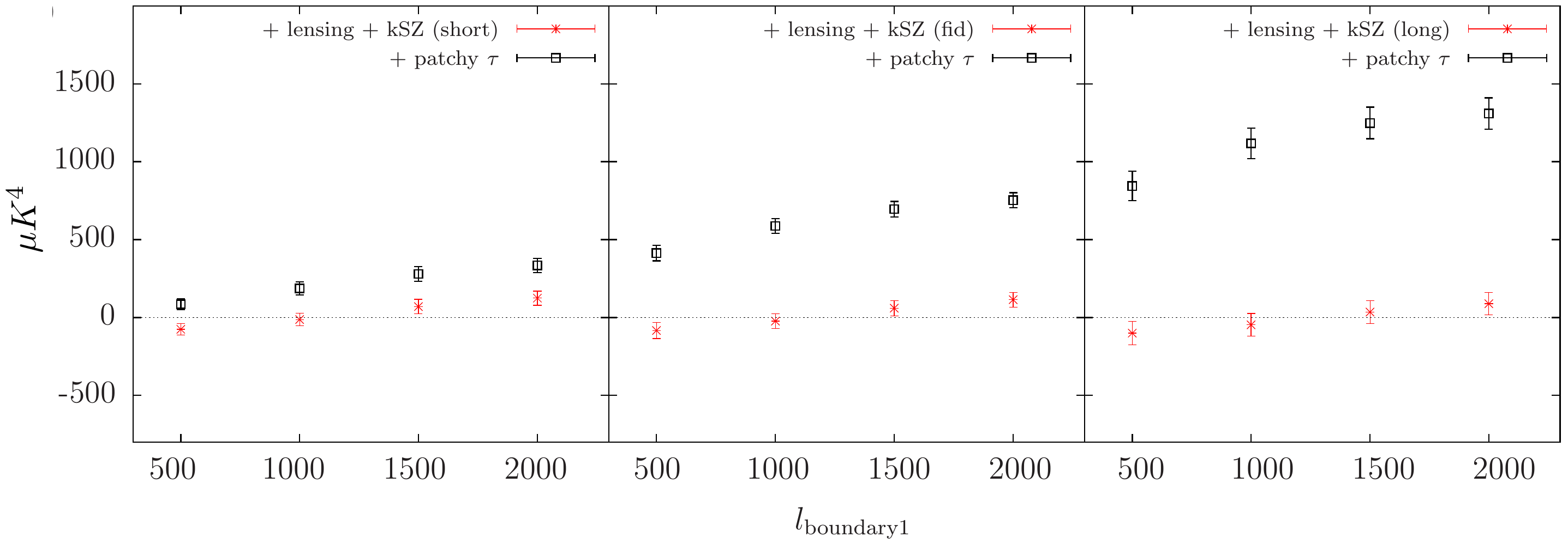}}
\end{center}
\caption{ 
The patchy $\tau$ correlator for different reionization scenarios. The CMB maps were obtained using {\scriptsize HEALPIX} \cite{healpix}, and the patchy $\tau$ maps were obtained from numerical simulations \cite{paper1}. Only kSZ and lensing have been included. It is easy to distinguish even small values of patchy $\tau$ when the larger secondary components have been removed through a multi-frequency analysis.
From Ref. \citen{paper2}. 
\label{arvi_fig7} }
\end{figure}

The observed fluctuation is then given by $\theta_{\rm obs}(\hat n) = \theta_{\rm cmb}(\hat n) \times \exp \left [-\delta\tau(\hat n) \right ] \approx \theta_{\rm cmb}(\hat n) - \delta\tau(\hat n) \theta_{\rm cmb}(\hat n)$, where we have dropped the constant term $\exp \left[ -\langle\tau\rangle \right ]$ because it is an overall multiplicative constant. The $\delta\tau(\hat n)$ fluctuations are on scales much smaller than the primary CMB fluctuations. When the observed CMB map is filtered, we obtain a large scale map $\theta_{\rm L}$, and a small scale map $\theta_{\rm S} + \theta_{\rm L} \delta\tau$.  The large scale and small scale modes of the CMB are independent of each other, and are therefore uncorrelated. The $\delta\tau(\hat n)$ field is also uncorrelated with the CMB fluctuations. The patchy $\tau$ correlator is therefore simply $\langle\delta f \delta g \rangle = \langle\delta\tau^2\rangle \times \left( \langle\theta^4_{\rm L}\rangle - \langle\theta^2_{\rm L}\rangle\right )$. The patchy $\tau$ correlator is thus sensitive to optical depth fluctuations and vanishes in the limit of homogeneous reionization.

Figure \ref{arvi_fig7} (from Ref. \citen{paper2}) shows the patchy $\tau$ correlator in units of $\mu K^4$, with and without patchy reionization. The CMB maps were simulated using the {\scriptsize HEALPix} software \cite{healpix}. The patchy reionization map was obtained using numerical simulations. The boundary for the large scale map $l_{\rm boundary1}$ is varied from 500 - 2000. The small scale map includes multipoles from 3000-5000. The maps shown in Figure \ref{arvi_fig7} include kSZ and lensing, but do not include tSZ, CIB, or radio contributions. Thus, we assume that the large frequency dependent contaminations may be removed through a multi-frequency analysis of the data.  The three panels are plotted for the 3 reionization scenarios shown in Figure \ref{arvi_fig6}. 

Since different multipoles of the primary CMB provide independent information, the correlation between large and small scale maps is zero for the primary CMB. Including the effect of CMB lensing however results in a non-zero cross correlation, because lensing of the CMB results in a redistribution of power, transferring CMB power from large scales to small scales. The cross correlation due to lensing is negative for small $l_{\rm boundary1}$, and increases, passing through zero at $l_{\rm boundary1} \sim 1200$. The patchy $\tau$ term is similarly correlated. The cross correlation due to patchy $\tau$ is always positive and increases with $l_{\rm boundary1}$. Moreover the patchy $\tau$ terms contributes significantly more to the cross correlation than the lensing term. Thus, one can detect patchy reionization at high significance by computing the cross correlation between the squared maps. It is however, much harder to measure the patchy $\tau$ correlator when other secondaries such as tSZ, CIB, and radio contributions are present \cite{paper2}.


\section{Future prospects}
A number of observational programs are aimed at detecting
the signatures of reionization in the CMB.
Data from the Planck mission will deliver accurate measurement of
the CMB polarization and the total Thomson optical depth,
from which 
details of  reionization 
can be derived \cite{Holder, liddle}. Ongoing experiments such as 
ACTPol \cite{actpol} and SPTPol \cite{sptpol} will be able to 
measure the kinematic Sunyaev-Zeldovich effect to higher accuracy.
The Cosmology Large Angular Scale Surveyor (CLASS) \cite{class} and the Primordial Inflation Polarization Explorer (PIPER) \cite{piper} are designed to measure the primordial $B$ mode polarization of the CMB, but they will also have the sensitivity to measure the $E$ mode to very high accuracy.  The CLASS instrument has a field of view of $19^\circ \times 14^\circ$ with a resolution of 1.5$^\circ$ FWHM, and will measure the polarization of the CMB at 40, 90, and 150 GHz from Cerro Toco in the Atacama desert of northern Chile.  PIPER is a balloon based experiment, and will fly in both the northern and southern hemispheres, achieving a sky coverage of 85\%. With 5120 detectors, PIPER is expected to obtain noise residuals less than 2.7 nK with 100 hours of observation. These experiments will significantly improve our understanding of the reionization history of the Universe through precise measurements of the large angle polarization of the CMB.

Future observations of the CMB spectral distortion
will open a new window to probe
the characteristic spectral energy distribution 
of the dominant sources of reionization, e.g. by the Primordial Inflation Explorer (PIXIE) \cite{pixie}, and the Polarized Radiation Imaging and Spectroscopy Mission (PRISM) \cite{prism}.  Heating of gas by early stars results in a Compton-$y$ distortion proportional to the temperature of the gas and the reionization optical depth. PIXIE is expected to measure the Compton distortion to an accuracy $y < 2 \times 10^{-9}$. In combination with PIXIE's measurement of the optical depth, PIXIE can determine the temperature of the intergalactic medium to 5\% precision at $z$=11 \cite{pixie}. In fact any form of energy injection
to the CMB can be studied to unprecedented accuracy,
and hence decay or annihilation of dark matter particles,
for example, 
can be inferred from the exact shape of the distortion
that encodes the time when dark matter 
decay/annihilation occurred \cite{Khatri}.

The abundances of light element atoms and ions such as
OI, NII, CII can be measured, in principle, 
by utilizing the characteristic CMB spectral (frequency-dependent) 
signatures and angular fluctuations generated 
by the resonant scattering and fine structure line 
emission.\cite{Basu, Hernandez}
Assuming that the first sources of light are also the 
first sources of the metals, one can trace the early
star-formation history from CMB observations.

Altogether, we have very good prospects that there will be significant progress
in the study of the Dark Ages in the next two decades, when 
next-generation radio telescope arrays and space-borne CMB
experiments probe the distribution 
of the intergalactic gas in the early universe.

\section*{Acknowledgements}
The present work is supported in part by the Grants-in-Aid
by the Ministry of Education, Science and Culture of Japan
(25287050: NY).
A.N. was funded by NASA grant NNX14AB57G. A.N. acknowledges partial financial support from the Pittsburgh Particle physics, Astrophysics, and Cosmology Center, and the Department of Physics and Astronomy at the University of Pittsburgh.
Portions of this research were conducted at the Jet 
Propulsion Laboratory, California Institute of Technology, 
which is supported by the National Aeronautics 
and Space Administration (NASA).

\end{document}